\documentclass[a4paper,12pt,english]{article}
\usepackage[english]{babel}
\usepackage{amsmath}
\usepackage{amsfonts}
\usepackage{amssymb}
\usepackage{geometry}
\usepackage{authblk}
\usepackage{enumitem}

\setlist{leftmargin=1.5cm}

\newcommand{\ie}{{\it i.e.,} }

\makeatletter
\def\@maketitle{%
\newpage
\null
\begin{flushright}
FIAN/TD/24-13
\end{flushright}
\vskip 2em%
\begin{center}%
\let \footnote \thanks
{\LARGE \@title \par}%
\vskip 1.5em%
{\large

\lineskip .5em%
\begin{tabular}[t]{c}%
\@author
\end{tabular}\par}
\vskip 1em%
{\large \@date}%
\end{center}
\par
\vskip 1.5em
}
\makeatother

\begin{document}

\thispagestyle{empty}
\title{\textbf{Gauge Non-Invariant Higher-Spin Currents in $4d$ Minkowski Space}}
\author{\textbf{P.A. Smirnov and M.A. Vasiliev}}
\affil{\small \emph{I.E.Tamm Department of Theoretical Physics, Lebedev Physical Institute of RAS,}\\
\emph{Leninsky prospect 53, 119991, Moscow, Russia}}
\date{}
\maketitle
\thispagestyle{empty}
\begin{abstract}
Conserved currents of any spin $t>0$ built from symmetric
massless gauge fields of any integer spin $s \geq t$ in {4d} Minkowski space
are found. In
particular,  stress-energy tensor for a higher-spin field of any
spin is constructed. Analogously to spin-two  stress
(pseudo)tensor, currents considered in this paper are not
gauge invariant. However, they are shown to generate gauge invariant
conserved charges. Besides expected parity even HS currents, we found
unexpected parity odd currents that generate less symmetries than the even ones.
It is argued that these odd currents unlikely admit a consistent $AdS$ deformation.
\end{abstract}

\newpage
\setcounter{page}{1}
\section{Introduction}
Conserved charges and currents play significant role in the field theory, 
being responsible for the field symmetries. We consider the further investigation of this problem 
is appropriate for a paper in honor of Andrei Alekseevich Slavnov, whose
wonderful scientific researches has been focused on different acpects of local and global
symmetries in quantum field theory 
(for example, \cite{Slavnov:1972fg, Slavnov:2002kg}).

Although conserved currents for higher-spin (HS) fields
were extensively studied in the literature
\cite{c2,c3,c4,c5,c6,c7,c8,Kaparulin:2011aa}, so far main attention was payed to the
gauge invariant currents. Generally a conserved current
carries three spins $(t, s_1, s_2)$, where $t$ is a spin of the
current, while $s_1$ and $s_2$ are spins of fields from which it is
constructed. In this paper, we  consider the case of  $s_1 = s_2 = s$.
For example, stress tensor ($t=2$) exists for matter
fields of arbitrary spins $s_1=s_2=s$. It is well known
however, that for $s=t=2$, the stress tensor is
not gauge invariant, corresponding to the so-called gravitational
stress pseudo-tensor \cite{LL}.
As shown by Deser and Waldron \cite{c6}, for $t=2$ analogous
 phenomenon occurs for all massless fields of spins $s>2$.
$t=1$ currents have similar property. The currents with $s < 1$
are gauge invariant (in this case, for a trivial reason since a spin-zero
field has no gauge symmetry), while the spin-one
current built from two massless spin-one fields is not.

The aim of this paper is to present
the full list of non-gauge invariant currents with
$s \geq t$ in $4d$ Minkowski space.
Being non-gauge invariant, such currents may
not immediately lead to consistent higher-spin (HS) Noether current interactions
(For the related discussion of HS interactions we refer
 the reader to \cite{Vasilev:2011xf,Joung:2013nma} and references therein.)
On the other hand, non-gauge invariant currents give rise to the
gauge invariant conserved charges.

Surprisingly, we found more conserved currents that was originally
anticipated. Namely, besides expected parity even HS currents we found
unexpected parity odd currents. The latter generate less global
symmetries than the even ones. As discussed in more detail in Conclusion,
 the odd currents can unlikely admit a consistent $AdS$ deformation and
 probably correspond to some currents of mixed symmetry fields in any
 dimension that amount to the symmetric fields at $d=4$.

The construction of currents and charges may be important for the
analysis of black-hole physics in  HS theory
\cite{Didenko:2009td,Iazeolla:2011cb}. To this end, however, the analysis
of this paper has to be generalized to $AdS$
background geometry since nonlinear HS gauge theory
can only be formulated in a curved background (for a review and
more references see \cite{c3}).
This problem will be considered elsewhere.

The paper is organized as follows. In Section 2, we recall
the frame-like formulation of massless fields in two-component spinor notation. In Section 3,
non-gauge invariant currents built from symmetric HS fields
in $4d$ Minkowski space are found and shown to give rise
to the gauge invariant conserved charges.
In Section 4, further perspectives are briefly discussed.

\subsection*{Conventions}
Greek indices $\mu, \nu, \rho, \lambda, \sigma$
are base and run from 0 to 3. Other Greek indices are spinorial and take values 1, 2.
The latter are raised and lowered by the \emph{sp}(2)
antisymmetric forms
$\varepsilon_{\alpha\beta}, \varepsilon^{\alpha\beta},
\varepsilon_{\dot\alpha\dot\beta}, \varepsilon^{\dot\alpha\dot\beta}$
in the following way
\begin{gather}
\varepsilon^{\alpha \beta}\varepsilon_{\alpha \gamma} = \delta^\beta_\gamma, \quad
\varepsilon^{\dot\alpha \dot\beta}\varepsilon_{\dot\alpha \dot\gamma} = \delta^{\dot\beta}_{\dot\gamma}, \\
A_\alpha = A^{\beta}\varepsilon_{\beta\alpha}, \quad
 A^\alpha = A_{\beta}\varepsilon^{\alpha\beta}, \quad
A_{\dot\alpha} = A^{\dot\beta}\varepsilon_{\dot\beta\dot\alpha}, \quad
A^{\dot\alpha} = A_{\dot\beta}\varepsilon^{\dot\alpha\dot\beta}.
\end{gather}

 Complex conjugation $\bar A$ relates dotted and undotted spinors.
Brackets ($[...]$) $\{...\}$  imply complete (anti)symmetrization, \ie
\begin{equation}
A_{[\alpha}B_{\beta]}=\frac{1}{2}(A_\alpha B_\beta-A_\beta B_\alpha)\,,\qquad
A_{\{\alpha}B_{\beta\}}=\frac{1}{2}(A_\alpha B_\beta+A_\beta B_\alpha).
\end{equation}
$A^{\alpha(m)}$  denotes a totally symmetric
multispinor  $A^{\{\alpha_1 \ldots \alpha_m\}}$.

The wedge symbol $\wedge$ is everywhere implicit.

\section{$4d$ massless HS fields in the frame-like approach}

In the metric-like formalism \cite{fronsdal},
an integer spin-$s$ symmetric massless field
is described  by a totally symmetric tensor
$\varphi_{\mu_{1}...\mu_{s}}$ subject to the
double tracelessness condition $\varphi_{\rho}{}^{\rho}{}_{\lambda}{}^{\lambda}{}_{\mu_5...\mu_s}=0$
which is nontrivial for $s\geq 4$.
The action is
\begin{multline}\label{action}
S[\varphi]=\frac{1}{2}\int d^4 x\bigl(\partial_\nu \varphi_{\mu_1...\mu_s}\partial^\nu\varphi^{\mu_1...\mu_s}
-\frac{1}{2}s(s-1)\partial_\nu\varphi^{\lambda}{}_{\lambda\mu_3...\mu_s}\partial^\nu\varphi_{\rho}{}^{\rho\mu_{3}...\mu_s}\\
+s(s-1)\partial_{\nu}\varphi^{\lambda}{}_{\lambda\mu_{3}...\mu_s}\partial_{\rho}\varphi^{\nu\rho\mu_3...\mu_s}
-s\partial_{\nu}\varphi^{\nu}{}_{\mu_{2}...\mu_s}\partial_{\rho}\varphi^{\rho\mu_2...\mu_s} \\
-\frac{1}{4}s(s-1)(s-2)\partial_{\nu}\varphi^{\nu\rho}{}_{\rho\mu_{4}...\mu_s}\partial_{\lambda}\varphi_{\sigma}{}^{\lambda\sigma\mu_{4}...\mu_s}\bigr)\,.
\end{multline}

Instead of deriving conserved currents from this action
using the Noether's theorem, that requires finding corresponding HS
symmetries, we will look for conserved currents using the
frame-like formulation where the HS metric-like field
is replaced by a frame field and a set of 1-form connections \cite{Vasiliev:1980as,2}
\begin{equation*}
s\geq2:\quad\varphi_{\mu_{1}...\mu_{s}}\rightarrow\{
\omega^{\alpha(m)}{}_{,}{}^{\dot\beta(n)}\mid m+n = 2(s-1)\}\,,
\quad \omega^{\alpha(m)}{}_{,}{}^{\dot\beta(n)}=dx^\mu \omega_{\mu}{}^{\alpha(m)}{}_{,}{}^{\dot\beta(n)},
\end{equation*}
which are symmetric in all dotted and all undotted spinor indices
and satisfy the reality condition \cite{2}
\begin{equation}
\omega_{\alpha(m),\dot\beta(n)}^\dag = \omega_{\beta(n),\dot\alpha(m)} .
\end{equation}
The frame-like field is a particular connection at $n=m=s-1$
\begin{equation}
h_{\mu}{}^{\alpha(s-1)}{}_{,}{}^{\dot\beta(s-1)}dx^{\mu}
:= \omega_{\mu}{}^{\alpha(s-1)}{}_{,}{}^{\dot\beta(s-1)}dx^{\mu}\,.
\end{equation}
By imposing appropriate constraints, the
connections $\omega^{\alpha(m),\dot\beta(n)}$ can be
expressed via $t = \frac{1}{2} |m-n|$ derivatives of the frame field \cite{2}.

$P$-reflection of the $x_2$ direction $(A_0,A_1,A_2,A_3) \rightarrow (A_0,A_1,-A_2,A_3)$
is defined as \cite{2}
\begin{gather}
P\omega_{\alpha(m),\dot\beta(n)}=\omega_{\beta(n),\dot\alpha(m)},\qquad
P \tilde h_{\alpha,\dot\beta}=\tilde h_{\beta,\dot\alpha}.
\end{gather}
Compared to Hermitian conjugation, the $P$-reflection does not affect
an order of product factors and a sign of the imaginary unit $\mathrm{i}$.

Background gravity is described by the vierbein
1-form $\tilde h^{\alpha}{}_{,}{}^{\dot\beta}$ and
1-form connections $\tilde\omega^{\dot\alpha\dot\beta}, \ \tilde\omega^{\alpha\beta}$.
Lorentz covariant derivative $\tilde D$ acts as usual
\begin{equation}
\tilde D A^{\alpha(m)}{}_{,}{}^{\dot\beta(n)}=dA^{\alpha(m)}{}_{,}{}^{\dot\beta(n)}+
m  \tilde \omega^{\alpha}{}_{\gamma}A^{\alpha(m-1)\gamma}{}_{,}{}^{\dot\beta(n)}
+n \tilde \omega^{\dot\beta}{}_{\dot\delta}A^{\alpha(m)}{}_{,}{}^{\dot\beta(n-1)\dot\delta}
\end{equation}
for any multispinor $A^{\alpha(m),\dot\beta(n)}$.
The torsion and curvature 2-forms are
\begin{gather}
\tilde R^{\alpha}{}_{,}{}^{\dot\beta}=\tilde D {\tilde h}^{\alpha}{}_{,}{}^{\dot\beta}
=d{\tilde h}^{\alpha}{}_{,}{}^{\dot\beta}+
{\tilde\omega}^{\alpha}{}_{\gamma}{\tilde h}^{\gamma}{}_{,}{}^{\dot\beta}+
{\tilde\omega}^{\dot\beta}{}_{\dot\delta}{\tilde h}^{\alpha}{}_{,}{}^{\dot\delta},\\
\tilde R^{\alpha\beta}=\tilde D{\tilde\omega}^{\alpha\beta}=d{\tilde\omega}^{\alpha\beta}+
2{\tilde\omega}^{\alpha}{}_{\gamma}{\tilde\omega}^{\beta\gamma}\,,\qquad
\tilde R^{\dot\alpha\dot\beta}=\tilde D{\tilde\omega}^{\dot\alpha\dot\beta}=d{\tilde\omega}^{\dot\alpha\dot\beta}+
2{\tilde\omega}^{\dot\alpha}{}_{\dot\gamma}{\tilde\omega}^{\dot\beta\dot\gamma}.
\end{gather}
$4d$ Minkowski space is described by the equations
\begin{equation}
\label{min}
\tilde R^{\alpha}{}_{,}{}^{\dot\beta}=0,\qquad
\tilde R^{\alpha\beta}=0, \qquad
\tilde R^{\dot\alpha\dot\beta}=0.
\end{equation}

Linearized HS curvatures are
\begin{align}
R^{\alpha(m)}{}_{,}{}^{\dot\beta(n)}=&\tilde D\omega^{\alpha(m)}{}_{,}{}^{\dot\beta(n)}+\nonumber\\
&+n\theta(m-n)\tilde h_{\gamma,}{}^{\dot\beta}\omega^{\gamma\alpha(m)}{}_{,}{}^{\dot\beta(n-1)}
+m\theta(n-m)\tilde h^{\alpha}{}_{,\dot\delta}\omega^{\alpha(m-1)}{}_{,}{}^{\dot\beta(n)\dot\delta},
\end{align}
where
\begin{gather}
\theta(x)=
\begin{cases}
1&\text{at $x \geq 0$;}\\
0&\text{at $x < 0$.}
\end{cases}
\end{gather}

Free field equations  for massless fields of spins $s\geq 2$
in Minkowski space
can be written in the form \cite{2}
\begin{align}
\label{EM}
&R^{\alpha(m)}{}_{,}{}^{\dot\beta(n)} = 0 \qquad \text{for} \qquad n>0, \ m>0, \ n+m = 2(s-1); \\
\label{EMCundot}
&R^{\alpha(m)} = C^{\alpha(m)\gamma\delta} \
\tilde h_{\gamma,\dot\theta} \ \tilde h_{\delta,}{}^{\dot\theta}
\qquad \text{for} \qquad m=2(s-1);\\
\label{EMCdot}
&R^{\dot\beta(n)} =
C^{\dot\beta(n)\dot\gamma\dot\delta} \ \tilde h_{\varphi,\dot\gamma} \ \tilde h^{\varphi}{}_{,\dot\delta}
\qquad \text{for} \qquad n=2(s-1).
\end{align}
They are equivalent to the equations of motion which
follow from the \mbox{action (\ref{action})}
supplemented with certain algebraic constraints which express
connections $\omega_{\alpha(m),\dot\beta(n)}$
via $\frac{1}{2}|m-n|$ derivatives of the dynamical frame-like HS field.
The multispinor \mbox{0-forms}
$C^{\alpha(2s)}$ and $C^{\dot\beta(2s)}$, which remain non-zero on-shell,
  are spin-$s$ analogues of the Weyl tensor in gravity.

HS gauge transformation is
\begin{multline}\label{gt}
\delta \omega^{\alpha(m)}{}_{,}{}^{\dot\beta(n)}
=\tilde D\epsilon^{\alpha(m)}{}_{,}{}^{\dot\beta(n)}\\
+n\theta(m-n)\tilde h_{\gamma,}{}^{\dot\beta}\epsilon^{\gamma\alpha(m)}{}_{,}{}^{\dot\beta(n-1)}
+m\theta(n-m)\tilde h^{\alpha}{}_{,\dot\delta}\epsilon^{\alpha(m-1)}{}_{,}{}^{\dot\beta(n)\dot\delta},
\end{multline}
where a gauge parameter $\epsilon^{\alpha(m)}{}_{,}{}^{\dot\beta(n)}(x)$ is an
arbitrary function of $x$.

\section{Currents in Minkowski space}

\subsection{Problem setting}

We will describe currents as Hodge dual differential
forms. In these terms, the on-shell closure condition for the latter
is traded for the current conservation condition. In this paper, we
consider currents built from two HS connections with $s_1 = s_2 = s$
in $d=4$ Minkowski space where a
spin-$t$ current is a $3$-form $J^{\alpha(t-1)}{}_{,}{}^{\dot\beta(t-1)}$.
Such a current will be shown to
contain $t$ derivatives of a spin-$s$ dynamical field for any $s$.
We consider currents with $1 \leq t \leq s$. Those with $t>s$
are built from the gauge invariant HS Weyl tensors and
their derivatives and are available in the literature \cite{c7,Kaparulin:2011aa}.

Conserved currents generate conserved charges. By the Noether theorem the
latter are generators of global symmetries. Hence, one
should expect as many conserved charges as global symmetry parameters.
The latter can be identified with those gauge symmetry parameters that leave
the gauge fields invariant. Denoting global symmetry parameters
$\xi_{\alpha(m),\dot\beta(n)}(x)$, from (\ref{gt}) it follows
that they have to obey the conditions
\begin{multline}\label{glob}
D \xi^{\alpha(m)}{}_{,}{}^{\dot\beta(n)}:=
\tilde D\xi^{\alpha(m)}{}_{,}{}^{\dot\beta(n)} \\
+n\theta(m-n)\tilde h_{\gamma,}{}^{ \dot\beta}\xi^{\gamma\alpha(m)}{}_{,}{}^{\dot\beta(n-1)}
+m\theta(n-m)\tilde h^{\alpha}{}_{,\dot\delta}\xi^{\alpha(m-1)}{}_{,}{}^{\dot\beta(n)\dot\delta}=0\,.
\end{multline}

For example, consider spin-two global symmetries with
parameters $\xi^{\alpha}{}_{,}{}^{\dot\beta}, \ \xi^{\alpha\alpha}, \ \xi^{\dot\beta\dot\beta}$
which obey
\begin{gather*}
\tilde D \xi^{\alpha}{}_{,}{}^{\dot\beta} + \tilde h_{\gamma,}{}^{\dot\beta} \xi^{\alpha\gamma}
+ \tilde h^{\alpha}{}_{,\dot\delta} \xi^{\dot\beta\dot\delta}=0, \quad
\tilde D \xi^{\alpha\alpha} = 0,\quad
\tilde D \xi^{\dot\beta\dot\beta} = 0.
\end{gather*}
In the Cartesian coordinate system this gives
\begin{gather*}
d \xi^{\alpha}{}_{,}{}^{\dot\beta} + \tilde h_{\gamma,}{}^{\dot\beta} \xi^{\alpha\gamma}
+ \tilde h^{\alpha}{}_{,\dot\delta} \xi^{\dot\beta\dot\delta}=0, \quad
d \xi^{\alpha\alpha} = 0,\quad
d \xi^{\dot\beta\dot\beta} = 0.
\end{gather*}
The general solution of this system is
\begin{gather*}
\xi^{\alpha}{}_{,}{}^{\dot\beta} = a^{\alpha}{}_{,}{}^{\dot\beta} -
x_{\gamma,}{}^{\dot\beta} b^{\alpha\gamma}
- x^{\alpha}{}_{,\dot\delta} c^{\dot\beta\dot\delta}\, , \quad
\xi^{\alpha\alpha}  = b^{\alpha\alpha} \, ,\quad
\xi^{\dot\beta\dot\beta}  = c^{\dot\beta\dot\beta} \, .
\end{gather*}
where  parameters $a^{\alpha}{}_{,}{}^{\dot\beta}, b^{\alpha\alpha}, c^{\dot\beta\dot\beta}$ are
arbitrary constants which parametrize Poincar\'e algebra.
Analogously, for a spin $s$, there is as many global symmetry parameters
as the gauge parameters $\epsilon_{\alpha(m),\dot\beta(n)}$ with $n+m=2(s-1)$.

To define a HS charge as an integral over a $3d$ space, we should find
such a current 3-form $J_{\alpha(m),\dot\beta(n)}(x)$
built from dynamical HS fields that
\begin{equation}\label{xiJ}
J _{\xi } (x)=\sum_{n,m}\xi_{\alpha(m),\dot\beta(n)}(x)
J^{\alpha(m)}{}_{,}{}^{\dot\beta(n)} (x), \quad n+m=2(t-1)
\end{equation}
is closed by virtue of (\ref{glob}) and the HS field equations (\ref{EM})-(\ref{EMCdot}).
This demands $J_{\alpha(m),\dot\beta(n)}(x)$ to obey the following
conservation equation
\begin{multline}\label{set}
D^* J^{\alpha(m)}{}_{,}{}^{\dot\beta(n)}:=
\tilde D J^{\alpha(m)}{}_{,}{}^{\dot\beta(n)} 
+(2t-n-1)\theta(n-m-1)\tilde h_{\gamma,}{}^{ \dot\beta}J^{\gamma\alpha(m)}{}_{,}{}^{\dot\beta(n-1)}\\
+(2t-m-1)\theta(m-n-1)\tilde h^{\alpha}{}_{,\dot\delta}J^{\alpha(m-1)}{}_{,}{}^{\dot\beta(n)\dot\delta} \simeq 0\, ,
\end{multline}
where $\simeq$ implies that the equality holds on-shell.
Note, that as a consequence of (\ref{set}) the spin-$t$ current
$J^{\alpha(t-1)}{}_{,}{}^{\dot\beta(t-1)}$
obeys the conventional conservation condition
\begin{equation}
\label{clo}
\tilde D J^{\alpha(t-1)}{}_{,}{}^{\dot\beta(t-1)} \simeq 0.
\end{equation}
Taking into account that 3-form currents are related to the usual
currents by the Levi-Civita symbol $\varepsilon^{\mu\nu\rho\sigma}$
which is parity-odd,
we  call the current $J_{\alpha(t-1),\dot\beta(t-1)}$  even if
$P J_{\alpha(t-1),\dot\beta(t-1)} = - J_{\beta(t-1),\dot\alpha(t-1)}$
and odd if $P J_{\alpha(t-1),\dot\beta(t-1)} =  J_{\beta(t-1),\dot\alpha(t-1)}$.

The derivatives $D$ and $D^*$ are defined in such a way that
\begin{equation}
\label{JDD}
d J _{\xi } (x)=\sum_{n,m}D\xi_{\alpha(m),\dot\beta(n)}(x)J^{\alpha(m)}{}_{,}{}^{\dot\beta(n)} (x)
+\sum_{n,m} \xi_{\alpha(m),\dot\beta(n)}(x)D^* J^{\alpha(m)}{}_{,}{}^{\dot\beta(n)}(x)\,.
\end{equation}
This means that $D$ and $D^*$ are mutually conjugated.
We call  $D$ and $D^*$ adjoint and co-adjoint derivatives, respectively,
since they result from the restriction of the respective derivatives of
the flat contraction of the HS algebra to its Poincar\'e subalgebra.

 Eq.~(\ref{JDD}) implies that the charge
\begin{equation}
\label{charge}
 Q_\xi = \int \limits_{M^3} J _\xi
\end{equation}
is conserved by virtue of (\ref{glob}) and (\ref{set}). As a result,
there are as many conserved charges $Q_\xi$ as independent global symmetry
parameters $\xi$.
Conservation of currents does not imply that they are invariant under the
gauge transformations (\ref{gt}). However, as shown below,
the gauge variation of $J_{\xi}$ is exact
\begin{equation}
\delta J_\xi(x) \simeq d H_\xi(x),
\end{equation}
so that the charge $Q_\xi$ turns out to be gauge invariant.

Nontrivial charges are represented by the current
$J_\xi(x)$ cohomology,
\ie closed currents modulo exact ones $J_\xi \simeq d \Psi_\xi$.
Since the currents should be closed on-shell,  \ie
by virtue of the free field equations (\ref{EM})-(\ref{EMCdot}),
analysis is greatly simplified by the fact that
all linearized HS curvatures $R^{\alpha(m)}{}_{,}{}^{\dot\beta(n)}$ with ${m>0, \ n>0}$
are zero on shell.

Thus, the problem is
\begin{enumerate}[label=(\alph*)]
\item to find all 3-forms $J^{\alpha(t-1)}{}_{,}{}^{\dot\beta(t-1)}$ built from the spin-$s$
fields with $1 \leq t \leq s$, which are
on-shell closed ({\it i.e.} obey (\ref{clo})) but not exact;
\item to find all  3-forms $J^{\alpha(m)}{}_{,}{}^{\dot\beta(n)}$,  $m \neq n$,  $m+n=2(t-1)$,
that obey (\ref{set});
\item to check that $J^{\alpha(t-1)}{}_{,}{}^{\dot\beta(t-1)}$ are Hermitian
$(J^{\alpha(t-1)}{}_{,}{}^{\dot\beta(t-1)})^\dag= J^{\beta(t-1)}{}_{,}{}^{\dot\alpha(t-1)}$;
\item to check that the HS charges are gauge invariant.
\end{enumerate}

Consider first some examples.

\subsection{Examples}
\subsubsection{Spin-one current}
The spin-one current $J$ carries no indices and generalizes the standard current
built from the complex scalar field
$j_{\mu}=\varphi\partial_{\mu}\varphi^*-\varphi^*\partial_{\mu}\varphi$
to massless fields of any integer spin containing
one derivative of the spin-$s$ dynamical field. As usual,
to construct nontrivial
spin-one current a color index is needed, \ie HS connections
$\omega^{i; \alpha(m)}{}_{,}{}^{\dot\beta(n)}$ are endowed with an additional index $i=1 \ldots N$
which labels independent dynamical fields. To contract color
indices, we introduce the real matrix $c_{ij}$  which can be either symmetric,
$c_{ij}=c_{ji}$, or antisymmetric, ${c_{ij}=-c_{ji}}$.

Consider a 3-form
\begin{multline}\label{J1}
J=c_{ij}  [\omega^{i; \varphi\gamma(s-2)}{}_{,}{}^{\dot\delta(s-1)}\omega^{j;}{}_{\gamma(s-2),\dot\delta(s-1)\dot\theta}
+\omega^{i; \varphi\gamma(s-1)}{}_{,}{}^{\dot\delta(s-2)}\omega^{j;}{}_{\gamma(s-1),\dot\delta(s-2)\dot\theta}
]\tilde h_{\varphi,}{}^{\dot\theta}.
\end{multline}

Note that the anticommutativity of the background vierbein 1-forms
\begin{equation}
\label{anticom}
\tilde h_{\alpha,}{}^{\dot\beta}\tilde h_{\gamma,}{}^{\dot\delta}=
-\tilde h_{\gamma,}{}^{\dot\delta}\tilde h_{\alpha,}{}^{\dot\beta}
\end{equation}
 implies that
\begin{equation}
\label{hhH}
\tilde h_{\alpha,}{}_{\dot\beta}\tilde h_{\gamma,}{}_{\dot\delta}=
\epsilon_{\alpha \gamma} \bar H_{\dot\beta\dot\delta}+
\epsilon_{\dot\beta\dot\delta} H_{\alpha \gamma}\,,
\end{equation}
where
\begin{equation}
\label{H}
H_{\alpha \beta} =\frac{1}{2}\tilde h_{\alpha, \dot{\gamma}} \tilde
h_{\beta,}{}^{ \dot{\gamma}}\,,\qquad \bar H_{\dot\alpha\dot\beta}=
\frac{1}{2}\tilde h_{\gamma, \dot{\alpha}} \tilde
h^\gamma{}_{,\dot{\beta}}
\end{equation}
provide a basis of 2-forms in $4d$ Minkowski space.

Using (\ref{hhH}) and (\ref{H}), the direct computation gives
\begin{multline}\label{J1comp}
d J=c_{ij}[R^{i; \varphi\gamma(s-2)}{}_{,}{}^{\dot\delta(s-1)}\omega^{j;}{}_{\gamma(s-2),\dot\delta(s-1)\dot\theta}
\tilde h_{\varphi,}{}^{\dot\theta}
-\omega^{i; \varphi\gamma(s-2)}{}_{,}{}^{\dot\delta(s-1)}R^{j;}{}_{\gamma(s-2),\dot\delta(s-1)\dot\theta}
\tilde h_{\varphi,}{}^{\dot\theta}\\
-R^{i; \varphi\gamma(s-1)}{}_{,}{}^{\dot\delta(s-2)}\omega^{j;}{}_{\gamma(s-1),\dot\delta(s-2)\dot\theta}
\tilde h_{\varphi,}{}^{\dot\theta}
-\omega^{i; \varphi\gamma(s-1)}{}_{,}{}^{\dot\delta(s-2)}R^{j;}{}_{\gamma(s-1),\dot\delta(s-2)\dot\theta}
\tilde h_{\varphi,}{}^{\dot\theta}\\
+s(\omega^{i; \gamma(s-2)}{}_{,}{}^{\dot\delta(s-1)\dot\psi}\omega^{j;}{}_{\gamma(s-2),\dot\delta(s-1)\dot\theta}
\bar H_{\dot\psi}{}^{\dot\theta}
+\omega^{i; \varphi\gamma(s-1)}{}_{,}{}^{\dot\delta(s-2)}\omega^{j;\chi}{}_{\gamma(s-1),\dot\delta(s-2)}
H_{\chi\varphi})\\
-(s-2)(\omega^{i; \varphi\gamma(s-3)}{}_{,}{}^{\dot\delta(s)}\omega^{j;}{}_{\chi\gamma(s-3),\dot\delta(s)}
H^{\chi}{}_{\varphi}
+\omega^{i; \gamma(s)}{}_{,}{}^{\dot\delta(s-3)\dot\psi}\omega^{j;}{}_{\gamma(s),\dot\delta(s-3)\dot\theta}
\bar H_{\dot\psi}{}^{\dot\theta})].
\end{multline}
For $s > 2$ the curvature-dependent terms vanish
by virtue of Eq.~(\ref{EM}). For $s=2$
$R^{\alpha(2)} = C^{\alpha(2)\gamma\delta} \ \tilde h_{\gamma,\dot\theta} \ \tilde h_{\delta,}{}^{\dot\theta}$
and
$
R^{\dot\beta(2)} = C^{\dot\beta(2)\dot\gamma\dot\delta} \
\tilde h_{\varphi,\dot\gamma} \ \tilde h^{\varphi}{}_{,\dot\delta}
$. In this case, the
$C$-dependent terms also vanish because antisymmetrization over any three
two-component indices gives zero. For example
\begin{equation}\label{X}
X:=\omega^{\varphi,\dot\delta}C_{\dot\delta\dot\chi\dot\theta\dot\psi}
\tilde h_{\gamma,}{}^{\dot\chi} \tilde h^{\gamma}{}_{,}{}^{\dot\psi} \tilde h_{\varphi,}{}^{\dot\theta}=0,
\end{equation}
because the antisymmetrization of three undotted spinor indices occurs due to
the anticomutativity
of the 1-forms $\tilde h_{\gamma, \dot\gamma}$.

As a result,  we obtain
\begin{multline}
d J \simeq c_{ij}[
s(\omega^{i; \gamma(s-2)}{}_{,}{}^{\dot\delta(s-1)\dot\psi}\omega^{j;}{}_{\gamma(s-2),\dot\delta(s-1)\dot\theta}
\bar H_{\dot\psi}{}^{\dot\theta}
+\omega^{i; \varphi\gamma(s-1)}{}_{,}{}^{\dot\delta(s-2)}\omega^{j;\chi}{}_{\gamma(s-1),\dot\delta(s-2)}
H_{\chi\varphi})\\
-(s-2)(\omega^{i; \varphi\gamma(s-3)}{}_{,}{}^{\dot\delta(s)}\omega^{j;}{}_{\chi\gamma(s-3),\dot\delta(s)}
H^{\chi}{}_{\varphi}
+\omega^{i; \gamma(s)}{}_{,}{}^{\dot\delta(s-3)\dot\psi}\omega^{j;}{}_{\gamma(s),\dot\delta(s-3)\dot\theta}
\bar H_{\dot\psi}{}^{\dot\theta})],
\end{multline}
which is zero if $c_{ij}$ is antisymmetric. $J$ is Hermitian, $J^\dag=J$,
if $c_{ij}$ is antisymmetric.

To prove that $J$ is not exact consider an on-shell improvement $d \Psi$ with a
2-form $\Psi$ containing no derivatives of the dynamical HS field
to give rise to a current with one derivative.
There is just one such form with \mbox{antisymmetric $c_{ij}$}
\begin{gather*}
\Psi = c_{ij}\omega^{i; \gamma(s-1)}{}_{,}{}^{\dot\delta(s-1)}\omega^{j;}{}_{\gamma(s-1),\dot\delta(s-1)} \, .
\end{gather*}
This gives
\begin{multline*}
d \Psi \simeq -2(s-1)c_{ij}(-\omega^{i; \varphi\gamma(s-2)}{}_{,}{}^{\dot\delta(s-1)}\omega^{j;}{}_{\gamma(s-2),\dot\delta(s-1)\dot\theta}
\tilde h_{\varphi,}{}^{\dot\theta}\\
+\omega^{i; \varphi\gamma(s-1)}{}_{,}{}^{\dot\delta(s-2)}\omega^{j;}{}_{\gamma(s-1),\dot\delta(s-2)\dot\theta}
\tilde h_{\varphi,}{}^{\dot\theta}).
\end{multline*}

Since the two terms in (\ref{J1}) have the same signs,
the 3-form $J$ is closed, but not exact, thus, representing the
on-shell current cohomology.

\subsubsection{Spin-two current}

The spin-two current is represented by the 3-form
\begin{equation}\label{J2}
J^{\alpha}{}_{,}{}^{\dot\beta}=2c_{ij}\omega^{i; \alpha\varphi\gamma(s-2)}{}_{,}{}^{\dot\delta(s-2)}
\omega^{j; }{}_{\gamma(s-2),\dot\delta(s-2)\dot\theta}{}^{\dot\beta}
\tilde h_{\varphi,}{}^{\dot\theta}\, ,
\end{equation}
which carries two derivatives of the HS field.
Like in the spin-one case, discarding the curvature-dependent terms
(for $s=2$, $C$-dependent terms vanish like in (\ref{X})),
the direct computation gives
\begin{equation}
\tilde D J^{\alpha}{}_{,}{}^{\dot\beta}\simeq 0
\end{equation}
regardless of the symmetry properties of $c_{ij}$.

To see that  the 3-form  $J^{\alpha}{}_{,}{}^{\dot\beta}$ is not exact
we observe that the only appropriate
2-forms are
\begin{align*}
\Psi_{1}{}^{\alpha}{}_{,}{}^{\dot\beta} &=c_{ij}\omega^{i; \alpha\gamma(s-2)}{}_{,}{}^{\dot\delta(s-1)}
\omega^{j;}{}_{\gamma(s-2),\dot\delta(s-1)}{}^{\dot\beta},\\
\Psi_{2}{}^{\alpha}{}_{,}{}^{\dot\beta} &=c_{ij}\omega^{i; \alpha\gamma(s-1)}{}_{,}{}^{\dot\delta(s-2)}
\omega^{j;}{}_{\gamma(s-1),\dot\delta(s-2)}{}^{\dot\beta}.
\end{align*}
An elementary computation gives
\begin{multline*}
\tilde D \Psi_{1}{}^{\alpha}{}_{,}{}^{\dot\beta} \simeq c_{ij}[
-\omega^{i; \gamma(s-2)}{}_{,}{}^{\dot\delta(s-1)\dot\theta}
\omega^{j;}{}_{\gamma(s-2),\dot\delta(s-1)}{}^{\dot\beta}\tilde h^{\alpha}{}_{,\dot\theta}\\
-(s-2)\omega^{i; \alpha\gamma(s-3)}{}_{,}{}^{\dot\delta(s-1)\dot\theta}
\omega^{j;}{}_{\varphi\gamma(s-3),\dot\delta(s-1)}{}^{\dot\beta}\tilde h^{\varphi}{}_{,\dot\theta}\\
-(s-1)\omega^{i; \alpha\varphi\gamma(s-2)}{}_{,}{}^{\dot\delta(s-2)}
-\omega^{j;}{}_{\gamma(s-2),\dot\delta(s-2)\dot\theta}{}^{\dot\beta}\tilde h_{\varphi,}{}^{\dot\theta}\\
-(s-2)\omega^{i; \alpha\varphi\gamma(s-3),\dot\delta(s-1)}
\omega^{j;}{}_{\gamma(s-3),\dot\delta(s-1)}{}^{\dot\theta\dot\beta}\tilde h_{\varphi,\dot\theta}],
\end{multline*}
\begin{multline*}
\tilde D \Psi_{2}{}^{\alpha}{}_{,}{}^{\dot\beta} \simeq c_{ij}[
-(s-2)\omega^{i; \alpha\varphi\gamma(s-1)}{}_{,}{}^{\dot\delta(s-3)}
\omega^{j;}{}_{\gamma(s-1),\dot\delta(s-3)\dot\theta}{}^{\dot\beta}\tilde h_{\varphi,}{}^{\dot\theta}\\
-\omega^{i; \alpha\gamma(s-1)}{}_{,}{}^{\dot\delta(s-2)}
\omega^{j;\varphi}{}_{\gamma(s-1),\dot\delta(s-2)}\tilde h_{\varphi,}{}^{\dot\beta}\\
-(s-2)\omega^{i; \alpha\gamma(s-1)}{}_{,}{}^{\dot\delta(s-3)\dot\theta}
\omega^{j;\varphi}{}_{\gamma(s-1),\dot\delta(s-3)}{}^{\dot\beta}\tilde h_{\varphi,\dot\theta}\\
-(s-1)\omega^{i; \alpha\varphi\gamma(s-2)}{}_{,}{}^{\dot\delta(s-2)}
\omega^{j;}{}_{\gamma(s-2),\dot\delta(s-2)}{}^{\dot\theta\dot\beta}\tilde h_{\varphi,\dot\theta}].
\end{multline*}
$J^{\alpha}{}_{,}{}^{\dot\beta}$ (\ref{J2}) is not a linear combination of
$\tilde D \Psi_{1}{}^{\alpha}{}_{,}{}^{\dot\beta}$ and
$\tilde D \Psi_{2}{}^{\alpha}{}_{,}{}^{\dot\beta}$ since the latter contain different
terms like
$c_{ij}\omega^{i; \alpha\gamma(s-1)}{}_{,}{}^{\dot\delta(s-2)}
\omega^{j;\varphi}{}_{\gamma(s-1),\dot\delta(s-2)}\tilde h_{\varphi,}{}^{\dot\beta}$
not present in (\ref{J2}). (Although by antisymmetrization of some three
two-component indices it is possible to reshuffle spinor indices
in this term, still $\tilde D \Psi_{1}{}^{\alpha}{}_{,}{}^{\dot\beta}$ and
$\tilde D \Psi_{2}{}^{\alpha}{}_{,}{}^{\dot\beta}$ contain extra terms compared to
(\ref{J2}).)

This leads to a surprising result that there are two independent
Hermitian conserved currents
(\ref{J2}) $J_{+}{}^{\alpha}{}_{,}{}^{\dot\beta} = J^{\alpha}{}_{,}{}^{\dot\beta}$ with $c_{ij}=c_{ji}$
and {$J_{-}{}^{\alpha}{}_{,}{}^{\dot\beta} =\mathrm{i} J^{\alpha}{}_{,}{}^{\dot\beta}$ with $ {c_{ij}=-c_{ji}}$.

Consider first the current $J_{+}^{\alpha}{}_{,}{}^{\dot\beta}$ which is even
because
$PJ_{+}{}^{\alpha}{}_{,}{}^{\dot\beta} = - J_{+}{}^{\beta}{}_{,}{}^{\dot\alpha}$.
For $J_{+}{}^{\alpha}{}_{,}{}^{\dot\beta}$ equations (\ref{set}) give
\begin{gather}
\label{eq1}
D^* J^{\dot\beta\dot\beta}_+ =
\tilde h_{\gamma,}{}^{\dot\beta}J_{+}{}^{\gamma}{}_{,}{}^{\dot\beta}
+ \tilde D J^{\dot\beta\dot\beta}_+ \simeq 0,\\
\label{eq2}
D^* J^{\alpha\alpha}_+ =
\tilde h^{\alpha}{}_{,\dot\delta}J_{+}{}^{\alpha}{}_{,}{}^{\dot\delta} +
\tilde D J^{\alpha\alpha}_+ \simeq 0\,.
\end{gather}
These equations are solved by
\begin{gather}
\label{Jbb}
J^{\dot\beta\dot\beta}_+=2 c_{ij}\omega^{i; \varphi\gamma(s-2)}{}_{,}{}^{\dot\delta(s-2)\dot\beta}
\omega^{j;}{}_{\gamma(s-2),\dot\delta(s-2)\dot\theta}{}^{\dot\beta}\tilde h_{\varphi,}{}^{\dot\theta},\\
\label{Jaa}
J^{\alpha\alpha}_+=2 c_{ij}\omega^{i; \alpha\varphi\gamma(s-2)}{}_{,}{}^{\dot\delta(s-2)}
\omega^{j;\alpha}{}_{\gamma(s-2),\dot\delta(s-2)\dot\theta}\tilde h_{\varphi,}{}^{\dot\theta}.
\end{gather}
Indeed, direct computation with the aid of (\ref{anticom}) gives on shell
\begin{multline*}
\tilde D J^{\dot\beta\dot\beta}_+ \simeq  2 c_{ij}[
s\omega^{i;\gamma(s-2)}{}_{,}{}^{\dot\delta(s-2)\dot\psi\dot\beta}
\omega^{j;}{}_{\gamma(s-2),\dot\delta(s-2)\dot\theta}{}^{\dot\beta}
\bar H_{\dot\psi}{}^{\dot\theta}\\
-(s-2)\omega^{i;\varphi\gamma(s-3)}{}_{,}{}^{\dot\delta(s-1)\dot\beta}
\omega^{j;}{}_{\chi\gamma(s-3),\dot\delta(s-1)}{}^{\dot\beta}
H^{\chi}{}_{\varphi}
-\omega^{i;\varphi\chi\gamma(s-2)}{}_{,}{}^{\dot\delta(s-2)}
\omega^{j;}{}_{\gamma(s-2),\dot\delta(s-2)\dot\theta}{}^{\dot\beta}
\tilde h_{\chi,}{}^{\dot\beta} \tilde h_{\varphi,}{}^{\dot\theta}
].
\end{multline*}
Here the first two terms vanish for symmetric $c_{ij}$.
As a result,
\begin{gather*}
\tilde D J^{\dot\beta\dot\beta}_+ \simeq -2 c_{ij}
\omega^{i;\varphi\chi\gamma(s-2)}{}_{,}{}^{\dot\delta(s-2)}
\omega^{j;}{}_{\gamma(s-2),\dot\delta(s-2)\dot\theta}{}^{\dot\beta}
\tilde h_{\chi,}{}^{\dot\beta} \tilde h_{\varphi,}{}^{\dot\theta}=
\tilde h_{\chi,}{}^{\dot\beta}J_{+}{}^{\chi}{}_{,}{}^{\dot\beta}.
\end{gather*}
Analogously, $J^{\alpha\alpha}_+$ (\ref{Jaa}) satisfies Eq.~(\ref{eq2}).

Thus, the even conserved current
$J_{+}{}^{\alpha}{}_{,}{}^{\dot\beta}$ (\ref{J2}) with
symmetric $c_{ij}$ gives rise to the closed current $J_\xi$ that generates
the full set of spin-two charges for  fields of any spin $s\geq 2$.
This current exists for $s \geq 2$ and corresponds to
the stress tensors for HS fields, which are the frame-like counterparts
of those considered in \cite{c6}.

Consider now the odd current $J_{-}{}^{\alpha}{}_{,}{}^{\dot\beta}$.
One can see that in this case Eqs.~(\ref{set}) admit no solutions.
As a result, the charge conservation condition demands
$\xi_{\alpha\alpha} = 0, \ \xi_{\dot\beta\dot\beta} = 0$.
The conserved charge is
\begin{equation}
Q_\xi=\int \limits_{M^3} \xi_{\alpha,\dot\beta}J_{-}{}^{\alpha}{}_{,}{}^{\dot\beta} \, ,
\end{equation}
where
$\xi_{\alpha,\dot\beta}$ satisfies
\begin{equation}\label{xi2}
\tilde D \xi_{\alpha,\dot\beta} = 0\,,
\end{equation}
hence being a constant in Cartesian coordinates.
Thus, the odd current $J_{-}{}^{\alpha}{}_{,}{}^{\dot\beta}$
generates less charges than  $J_{+}{}^{\alpha}{}_{,}{}^{\dot\beta}$.

For example, in tensor notations, the odd current for $s=t=2$ reads as
\begin{equation*}
J^a = c_{ij}[\omega^{i; mn}\omega^{j;}{}_{mn}\tilde e^a -
4 \omega^{i; an}\omega^{j;}{}_{mn}\tilde e^m ],
\end{equation*}
where $c_{ij}$ is antisymmetric.

The existence of odd conserved currents is surprising. Analogous
``unexpected" currents are shown in the next section to exist for
all higher spin of currents with $t>2$ in Minkowski space.
In Conclusion we argue that the odd currents
only exist in Minkowski space and unlikely admit an extension to
$AdS_4$.

\subsection{General case}
For $t \geq 1$, a conserved current
$J^{\alpha(t-1)}{}_{,}{}^{\dot\beta(t-1)}$ for $s \geq 2, s \geq t$,
that carries $t$ derivatives of the HS field, is
\begin{multline}\label{gen}
J^{\alpha(t-1)}{}_{,}{}^{\dot\beta(t-1)} = c_{ij}
 [\omega^{i; \alpha(t-1)\varphi\gamma(s-2)}{}_{,}{}^{\dot\delta(s-t)}
\omega^{j;}{}_{\gamma(s-2),\dot\delta(s-t)\dot\theta}{}^{\dot\beta(t-1)}
\tilde h_{\varphi,}{}^{\dot\theta} \\
+ \omega^{i; \alpha(t-1)\varphi\gamma(s-t)}{}_{,}{}^{\dot\delta(s-2)}
\omega^{j;}{}_{\gamma(s-t),\dot\delta(s-2)\dot\theta}{}^{\dot\beta(t-1)}
\tilde h_{\varphi,}{}^{\dot\theta}].
\end{multline}
Analogously to the case of $t=2$, it is not difficult to see that
$\tilde D J^{\alpha(t-1)}{}_{,}{}^{\dot\beta(t-1)} \simeq 0$ regardless of
the symmetry of $c_{ij}$. For $t=s-1$, the terms with
Weyl-like tensors vanish like in the example (\ref{X}).
The proof of the fact that the current $J^{\alpha(t-1)}{}_{,}{}^{\dot\beta(t-1)}$ (\ref{gen}) is not exact
regardless of the symmetry of $c_{ij}$
is straightforward but technically involved. The
details are given in Appendix.

Analogously to the case of $t=2$
(\ref{J2}) there are two different Hermitian conserved currents
$J_{+}{}^{\alpha(t-1)}{}_{,}{}^{\dot\beta(t-1)} = J^{\alpha(t-1)}{}_{,}{}^{\dot\beta(t-1)}$, where
$c_{ij}$ is (anti)symmetric for (odd)even $t$, and
$J_{-}{}^{\alpha(t-1)}{}_{,}{}^{\dot\beta(t-1)} = \mathrm{i} J^{\alpha(t-1)}{}_{,}{}^{\dot\beta(t-1)}$,
where $c_{ij}$ is (anti)symmetric for (even)odd $t$.
Note, that  Eq.~(\ref{gen}) at $t=2$ gives  $J^{\alpha,\dot\beta}_+$ (\ref{J2}).

Consider the even current $J_{+}{}^{\alpha(t-1)}{}_{,}{}^{\dot\beta(t-1)}$.
To reconstruct the current $J_\xi$ we have to solve
the system of equations (\ref{set}).
Using equations (\ref{EM}) - (\ref{EMCdot}) one can see
that the following set of 3-forms
$J_{+}{}^{\alpha(m)}{}_{,}{}^{\dot\beta(n)}, m \neq n, m+n=2(t-1)$ gives a solution
\begin{multline}\label{aux}
J_{+}{}^{\alpha(m)}{}_{,}{}^{\dot\beta(n)}
=\theta(n-m-4) \ g(n) \ c_{ij}\omega^{i;\alpha(m)\varphi\gamma(s-2)}{}_{,}{}^{\dot\delta(s-t)\dot\beta(n-t+1)}
\omega^{j;}{}_{\gamma(s-2),\dot\delta(s-t)\dot\theta}{}^{\dot\beta(t-1)}
\tilde h_{\varphi,}{}^{\dot\theta}\\
+\delta_{n,t} \ c_{ij}[2(t-1)\omega^{i;\alpha(m)\varphi\gamma(s-2)}{}_{,}{}^{\dot\delta(s-t)\dot\beta}
\omega^{j;}{}_{\gamma(s-2),\dot\delta(s-t)\dot\theta}{}^{\dot\beta(n-1)} \tilde h_{\varphi,}{}^{\dot\theta}\\
+\sum \limits_{p=1}^{t-2} f(p)\omega^{i;\alpha(m)\varphi\gamma(s-p-1)}{}_{,}{}^{\dot\delta(s-t+p)}
\omega^{j;}{}_{\gamma(s-p-1),\dot\delta(s-t+p)}{}^{\dot\beta(n-1)} \tilde h_{\varphi,}{}^{\dot\beta}]\\
+\theta(m-n-4) \ g(m) \ c_{ij}\omega^{i;\alpha(t-1)\varphi\gamma(s-t)}{}_{,}{}^{\dot\delta(s-2)}
\omega^{j;\alpha(m-t+1)}{}_{\gamma(s-t),\dot\delta(s-2)\dot\theta}{}^{\dot\beta(n)}
 \tilde h_{\varphi,}{}^{\dot\theta}\\
+\delta_{m,t} \ c_{ij}[2(t-1)\omega^{i;\alpha(m-1)\varphi\gamma(s-t)}{}_{,}{}^{\dot\delta(s-2)}
\omega^{j;\alpha}{}_{\gamma(s-t),\dot\delta(s-2)\dot\theta}{}^{\dot\beta(n)} \tilde h_{\varphi,}{}^{\dot\theta}\\
+\sum \limits_{p=1}^{t-2} f(p)\omega^{i;\alpha(m-1)\gamma(s-t+p)}{}_{,}{}^{\dot\delta(s-p-1)}
\omega^{j;}{}_{\gamma(s-t+p),\dot\delta(s-p-1)}{}^{\dot\theta\dot\beta(n)} \tilde h^{\alpha}{}_{,\dot\theta}],
\end{multline}
where
\begin{gather}
g(m) =\frac{2(t-1)!}{(2t-m-2)!(m-t+1)!}\,, \quad m \geq t+1,\\
f(1) = \frac{t - 1}{s-t+1}\,, \quad f(p)=(t-1)\frac{(s-t)!(s-p)!}{(s-3)!(s-t+p)!}\,, \quad p>1.
\end{gather}
It should be stressed, that the forms $J_{+}^{\dot\beta(2t-2)}$,
$J_{+}^{\alpha(2t-2)}$
obey equations
(\ref{set}) only if
$c_{ij}$ is antisymmetric for odd $t$ and symmetric for even $t$.

Thus, the Hermitian current 3-form $J_{+}{}^{\alpha(t-1)}{}_{,}{}^{\dot\beta(t-1)}$ (\ref{gen})
with antisymmetric $c_{ij}$ for odd $t$ and symmetric for even $t$
is on-shell closed, but not exact. It generates the closed current
$J_\xi$ and  the corresponding conserved charge
$Q_\xi=\int \limits_{M^3} J_\xi$ that contains as many symmetry parameters as
local HS gauge symmetries.

Since the current $J^{\alpha(t-1)}{}_{,}{}^{\dot\beta(t-1)}$ (\ref{gen}) is closed
but not exact regardless of the symmetry of $c_{ij}$, the odd Hermitian current
with symmetric $c_{ij}$ for odd $t$ and antisymmetric for even is
${J_{-}{}^{\alpha(t-1)}{}_{,}{}^{\dot\beta(t-1)} =
\mathrm{i} J^{\alpha(t-1)}{}_{,}{}^{\dot\beta(t-1)}}$.
Odd currents exist for $t \geq 2$.
To reconstruct $J_\xi$ we have to solve equations (\ref{set}).
It turns out that these equations is possible to solve at $m \neq 2(t-1), n \neq 2(t-1), m+n=2(t-1)$.
The solution is expressed by
$J_{-}{}^{\alpha(m)}{}_{,}{}^{\dot\beta(n)}=
\mathrm{i}J_{+}{}^{\alpha(m)}{}_{,}{}^{\dot\beta(n)}$
(\ref{aux}), but $c_{ij}$ is antisymmetric for even $t$ and symmetric for odd $t$.
However the equations admit no solution at the last step, i.e., for
$J_{-}^{\dot\beta(2t-2)}$
and $J_{-}^{\alpha(2t-2)}$.
Hence, as in the example of $t=2$, the charge conservation condition
demands $\xi_{\alpha(2t-2)}=0$, $\xi_{\dot\beta(2t-2)}=0$.
Since Eq.~(\ref{glob}) admits less solutions,
 the odd currents generate less charges than the even ones.

\subsection{Gauge transformations}
The on-shell gauge variation of the current 3-form (\ref{gen})
can be represented in the form
\begin{multline}
\delta J^{\alpha(t-1)}{}_{,}{}^{\dot\beta(t-1)} \simeq c_{ij}  \tilde D [
 \epsilon^{i; \alpha(t-1)\varphi\gamma(s-2)}{}_{,}{}^{\dot\delta(s-t)}
\omega^{j;}{}_{\gamma(s-2),\dot\delta(s-t)\dot\theta}{}^{\dot\beta(t-1)}
\tilde h_{\varphi,}{}^{\dot\theta}\\
+\omega^{i; \alpha(t-1)\varphi\gamma(s-2)}{}_{,}{}^{\dot\delta(s-t)}
\epsilon^{j;}{}_{\gamma(s-2),\dot\delta(s-t)\dot\theta}{}^{\dot\beta(t-1)}
\tilde h_{\varphi,}{}^{\dot\theta}\\
+ \epsilon^{i; \alpha(t-1)\varphi\gamma(s-t)}{}_{,}{}^{\dot\delta(s-2)}
\omega^{j;}{}_{\gamma(s-t),\dot\delta(s-2)\dot\theta}{}^{\dot\beta(t-1)}
\tilde h_{\varphi,}{}^{\dot\theta}  \\
+ \omega^{i; \alpha(t-1)\varphi\gamma(s-t)}{}_{,}{}^{\dot\delta(s-2)}
\epsilon^{j;}{}_{\gamma(s-t),\dot\delta(s-2)\dot\theta}{}^{\dot\beta(t-1)}
\tilde h_{\varphi,}{}^{\dot\theta}] = \tilde D H^{\alpha(t-1)}{}_{,}{}^{\dot\beta(t-1)}.
\end{multline}
Analogously, the on-shell gauge variation of the 3-forms (\ref{aux}) is
\begin{gather}
\delta J_{+}{}^{\alpha(m)}{}_{,}{}^{\dot\beta(n)}
\simeq D^* H^{\alpha(m)}{}_{,}{}^{\dot\beta(n)},
\quad m \neq n,  \\
\delta J_{-}{}^{\alpha(m)}{}_{,}{}^{\dot\beta(n)}
\simeq \mathrm{i}D^* H^{\alpha(m)}{}_{,}{}^{\dot\beta(n)}, \quad
m \neq n, \quad m \neq 2(t-1), \quad n \neq 2(t-1)\,.
\end{gather}
The 2-form $H^{\alpha(m)}{}_{,}{}^{\dot\beta(n)}$ is
\begin{multline}
H^{\alpha(m)}{}_{,}{}^{\dot\beta(n)}
=\theta(n-m-4) \ g(n) \ c_{ij}[\epsilon^{i;\alpha(m)\varphi\gamma(s-2)}{}_{,}{}^{\dot\delta(s-t)\dot\beta(n-t+1)}
\omega^{j;}{}_{\gamma(s-2),\dot\delta(s-t)\dot\theta}{}^{\dot\beta(t-1)}
\tilde h_{\varphi,}{}^{\dot\theta}\\
+\omega^{i;\alpha(m)\varphi\gamma(s-2)}{}_{,}{}^{\dot\delta(s-t)\dot\beta(n-t+1)}
\epsilon^{j;}{}_{\gamma(s-2),\dot\delta(s-t)\dot\theta}{}^{\dot\beta(t-1)}
\tilde h_{\varphi,}{}^{\dot\theta}]\\
+\delta_{n,t} \ c_{ij}[2(t-1)\epsilon^{i;\alpha(m)\varphi\gamma(s-2)}{}_{,}{}^{\dot\delta(s-t)\dot\beta}
\omega^{j;}{}_{\gamma(s-2),\dot\delta(s-t)\dot\theta}{}^{\dot\beta(n-1)} \tilde h_{\varphi,}{}^{\dot\theta}\\
+2(t-1)\omega^{i;\alpha(m)\varphi\gamma(s-2)}{}_{,}{}^{\dot\delta(s-t)\dot\beta}
\epsilon^{j;}{}_{\gamma(s-2),\dot\delta(s-t)\dot\theta}{}^{\dot\beta(n-1)} \tilde h_{\varphi,}{}^{\dot\theta}\\
+\sum \limits_{p=1}^{t-2} f(p)\epsilon^{i;\alpha(m)\varphi\gamma(s-p-1)}{}_{,}{}^{\dot\delta(s-t+p)}
\omega^{j;}{}_{\gamma(s-p-1),\dot\delta(s-t+p)}{}^{\dot\beta(n-1)} \tilde h_{\varphi,}{}^{\dot\beta}\\
+\sum \limits_{p=1}^{t-2} f(p)\omega^{i;\alpha(m)\varphi\gamma(s-p-1)}{}_{,}{}^{\dot\delta(s-t+p)}
\epsilon^{j;}{}_{\gamma(s-p-1),\dot\delta(s-t+p)}{}^{\dot\beta(n-1)} \tilde h_{\varphi,}{}^{\dot\beta}]\\
+\theta(m-n-4) \ g(m) \ c_{ij}[\epsilon^{i;\alpha(t-1)\varphi\gamma(s-t)}{}_{,}{}^{\dot\delta(s-2)}
\omega^{j;\alpha(m-t+1)}{}_{\gamma(s-t),\dot\delta(s-2)\dot\theta}{}^{\dot\beta(n)}
 \tilde h_{\varphi,}{}^{\dot\theta}\\
+\omega^{i;\alpha(t-1)\varphi\gamma(s-t)}{}_{,}{}^{\dot\delta(s-2)}
\epsilon^{j;\alpha(m-t+1)}{}_{\gamma(s-t),\dot\delta(s-2)\dot\theta}{}^{\dot\beta(n)}
 \tilde h_{\varphi,}{}^{\dot\theta}]\\
+\delta_{m,t} \ c_{ij}[2(t-1)\epsilon^{i;\alpha(m-1)\varphi\gamma(s-t)}{}_{,}{}^{\dot\delta(s-2)}
\omega^{j;\alpha}{}_{\gamma(s-t),\dot\delta(s-2)\dot\theta}{}^{\dot\beta(n)} \tilde h_{\varphi,}{}^{\dot\theta}\\
+2(t-1)\omega^{i;\alpha(m-1)\varphi\gamma(s-t)}{}_{,}{}^{\dot\delta(s-2)}
\epsilon^{j;\alpha}{}_{\gamma(s-t),\dot\delta(s-2)\dot\theta}{}^{\dot\beta(n)} \tilde h_{\varphi,}{}^{\dot\theta}\\
+\sum \limits_{p=1}^{t-2} f(p)\epsilon^{i;\alpha(m-1)\gamma(s-t+p)}{}_{,}{}^{\dot\delta(s-p-1)}
\omega^{j;}{}_{\gamma(s-t+p),\dot\delta(s-p-1)}{}^{\dot\theta\dot\beta(n)} \tilde h^{\alpha}{}_{,\dot\theta}\\
+\sum \limits_{p=1}^{t-2} f(p)\omega^{i;\alpha(m-1)\gamma(s-t+p)}{}_{,}{}^{\dot\delta(s-p-1)}
\epsilon^{j;}{}_{\gamma(s-t+p),\dot\delta(s-p-1)}{}^{\dot\theta\dot\beta(n)} \tilde h^{\alpha}{}_{,\dot\theta}],
\end{multline}
where $m+n=2(t-1)$.

As a result, the gauge transformation of  $Q_\xi$ is
\begin{multline}
\delta Q_\xi
\simeq
\int \limits_{M^3}  \sum \limits_{m,n}  \xi_{\alpha(m),\dot\beta(n)}  D^* H^{\alpha(m)}{}_{,}{}^{\dot\beta(n)}  \\
= \int \limits_{M^3} d \Big( \sum \limits_{m,n}
\xi_{\alpha(m),\dot\beta(n)}H^{\alpha(m)}{}_{,}{}^{\dot\beta(n)}\Big) = \int \limits_{M^3} d H_\xi =0.
\end{multline}
Thus, though the current $J_\xi$ is not gauge invariant,
the corresponding charge is.

\section{Conclusion}
In this paper, spin-$t$ HS even currents $J_{+}{}^{\alpha(t-1)}{}_{,}{}^{\dot\beta(t-1)}$ in
four-dimensional Minkowski space,
built from spin-$s$ $(s \geq t)$ fields and carrying $t$ derivatives
(which is the minimal possible number), are found.
Being represented as 3-forms, $J^{\alpha(t-1)}{}_{,}{}^{\dot\beta(t-1)}$
are closed but not exact, hence leading to nontrivial HS charges.
To generate the full list of conserved charges,
 $J^{\alpha(t-1)}{}_{,}{}^{\dot\beta(t-1)}$ (\ref{gen}) was extended to a set of currents
 $J^{\alpha(m)}{}_{,}{}^{\dot\beta(n)}$ with $m+n=2(t-1)$ that altogether obey
 the co-adjoint covariant constancy condition and form a full current $J_\xi$
 that depends on the global symmetry parameters $\xi$.

Analogously to the case of stress tensor ($t=2$)
considered in \cite{c6}, the constructed currents are not invariant under the
HS gauge transformations. However, the corresponding HS charges are gauge invariant
because the gauge variation $\delta J_\xi$ is exact.

In addition to the expected set of $P$-even currents related to
usual HS symmetries we found an unexpected set of spin $t>1$ $P$-odd currents
which have opposite symmetry with respect to the color indices carried by
HS fields compared to the normal  $P$-even currents.
The origin of these currents is rather mysterious. We expect that
beyond four dimensions they should correspond to currents of
some mixed symmetry fields that become equivalent to symmetric fields in
the particular case of four dimensions. A related property is that the
odd currents can unlikely survive upon the deformation of Minkowski space
to $AdS_4$. This  follows from the fact that, as shown in this paper,
 the space of global symmetry  parameters associated with the odd currents
 is smaller than for even currents. Such a reduction is possible in Minkowski
 geometry where HS connections are valued in a indecomposable module
 of the Poincar\'e algebra allowing a reduction to a submodule. However, it cannot
 be possible in the $AdS$ geometry where the $AdS_4$ symmetry $sp(4)$ is
 simple and HS connections are valued in its irreducible modules for any
 given spin. In fact, this argument fits the property that mixed symmetry
 fields in (anti-)de Sitter and Minkowski spaces are essentially different
 \cite{Metsaev:1995re,Brink:2000ag} in the sense that $(A)dS$ mixed symmetry fields
 contain more degrees of freedom than the Minkowski ones that obstructs
 a smooth $(A)dS$ deformation of the currents associated with mixed symmetry fields.

It would be interesting to generalize obtained results to currents
built from fields of different spins and/or of the connection-Weyl (\ie
$\omega \times C$) type. For example, in the case of $t=s=1$,
the current is
\begin{equation*}
J =\mathrm{i} c_{ij} [\omega^{i} C^{j; \alpha\alpha}H_{\alpha\alpha} -
\omega^i C^{j;\dot\beta\dot\beta}\bar H_{\dot\beta\dot\beta}],
\end{equation*}
where $c_{ij}$ is antisymmetric.

To make it possible to use HS charges for
the analysis of particular solutions
of nonlinear HS gauge theory, it is important to generalize the constructed
currents to $(A)dS$ background, which problem is currently under investigation.
As mentioned above, it is also interesting to clarify the fate of odd HS
currents in $AdS_4$.

\section*{Acknowledgements}
This research was supported by RFBR Grant No 11-02-00814-a and
RFBR Grant No 12-02-31838.

\section*{Appendix. Non-exactness of $J^{\alpha(t-1),\dot\beta(t-1)}$}
To check whether or not the current (\ref{gen}) can be represented in the
exact form we set
\begin{equation}
\label{phiphi}
J^{\alpha(t-1)}{}_{,}{}^{\dot\beta(t-1)} =
\Phi_{1}{}^{\alpha(t-1)}{}_{,}{}^{\dot\beta(t-1)} + \Phi_{2}{}^{\alpha(t-1)}{}_{,}{}^{\dot\beta(t-1)}.
\end{equation}
\begin{gather*}
\Phi_{1}{}^{\alpha(t-1)}{}_{,}{}^{\dot\beta(t-1)} =
c_{ij} \omega^{i; \alpha(t-1)\varphi\gamma(s-2)}{}_{,}{}^{\dot\delta(s-t)}
\omega^{j;}{}_{\gamma(s-2),\dot\delta(s-t)\dot\theta}{}^{\dot\beta(t-1)}
\tilde h_{\varphi,}{}^{\dot\theta}, \\
\Phi_{2}{}^{\alpha(t-1)}{}_{,}{}^{\dot\beta(t-1)} =
c_{ij} \omega^{i; \alpha(t-1)\varphi\gamma(s-t)}{}_{,}{}^{\dot\delta(s-2)}
\omega^{j;}{}_{\gamma(s-t),\dot\delta(s-2)\dot\theta}{}^{\dot\beta(t-1)}
\tilde h_{\varphi,}{}^{\dot\theta},
\end{gather*}

A basis in the space of 2-forms,
which are bilinear in fields,  is
\begin{align*}
\Psi^{\alpha(t-1)}{}_{,}{}^{\dot\beta(t-1)} (k, l)
= c_{ij} \omega^{i; \alpha(t-l-1)\gamma(s-k)}{}_{,}{}^{\dot\delta(s-t+k-1)\dot\beta(l)}
\omega^{j;\alpha(l)}{}_{\gamma(s-k),\dot\delta(s-t+k-1)}{}^{\dot\beta(t-l-1)}.
\end{align*}
The choice of $k, l$ is restricted by the condition
 that $\Psi^{\alpha(t-1)}{}_{,}{}^{\dot\beta(t-1)} (k, l)$
contains $(t-1)$ derivatives of the dynamical field.

A current is exact if
$J^{\alpha(t-1)}{}_{,}{}^{\dot\beta(t-1)}
\simeq \sum \limits_{k,l} a_{kl} \tilde D \Psi^{\alpha(t-1)}{}_{,}{}^{\dot\beta(t-1)} (k,l)$.
It is not difficult to check, that the forms
$\tilde D \Psi^{\alpha(t-1)}{}_{,}{}^{\dot\beta(t-1)} (k, l)$ with
$l \neq 0$ cannot contribute
because they give rise to non-vanishing terms different from
$\Phi_{1,2}{}^{\alpha(t-1)}{}_{,}{}^{\dot\beta(t-1)}$.
Thus, the current (\ref{phiphi}) should necessarily
contain $\tilde D \Psi^{\alpha(t-1)}{}_{,}{}^{\dot\beta(t-1)} (2, 0)$
and ${\tilde D \Psi^{\alpha(t-1)}{}_{,}{}^{\dot\beta(t-1)} (t-1, 0)}$
because among the forms with $l = 0$ only these contain
$\Phi_{1,2}{}^{\alpha(t-1)}{}_{,}{}^{\dot\beta(t-1)}$.

Let us try to construct the possible improvement
(coefficients are not important and denoted as $a_i > 0$)
\begin{equation}
I^{\alpha(t-1)}{}_{,}{}^{\dot\beta(t-1)} = -a_1 \tilde D \Psi^{\alpha(t-1)}{}_{,}{}^{\dot\beta(t-1)} (2, 0) + \ldots\,.
\end{equation}
The on-shell differential of $\Psi^{\alpha(t-1),\dot\beta(t-1)} (2, 0)$ is
\begin{multline*}
-\tilde D \Psi^{\alpha(t-1)}{}_{,}{}^{\dot\beta(t-1)} (2, 0) \simeq \\
\simeq (s-t+1)\Phi_{1}{}^{\alpha(t-1)}{}_{,}{}^{\dot\beta(t-1)}
-(s-2)c_{ij}\omega^{i; \alpha(t-1)\varphi\gamma(s-3)}{}_{,}{}^{\dot\delta(s-t+1)}
\omega^{j;}{}_{\gamma(s-3),\dot\delta(s-t+1)\dot\psi}{}^{\dot\beta(t-1)} \tilde h_{\varphi,}{}^{\dot\psi}].
\end{multline*}
The second term can be  subtracted by
\begin{multline*}
\tilde D \Psi^{\alpha(t-1)}{}_{,}{}^{\dot\beta(t-1)} (3, 0) \simeq \\
\simeq c_{ij} [-(s-t+2)\omega^{i; \alpha(t-1)\varphi\gamma(s-3)}{}_{,}{}^{\dot\delta(s-t+1)}
\omega^{j;}{}_{\gamma(s-3),\dot\delta(s-t+1)\dot\psi}{}^{\dot\beta(t-1)} \tilde h_{\varphi,}{}^{\dot\psi}+\\
+(s-3)\omega^{i; \alpha(t-1)\varphi\gamma(s-4)}{}_{,}{}^{\dot\delta(s-t+2)}
\omega^{j;}{}_{\gamma(s-4),\dot\delta(s-t+2)\dot\psi}{}^{\dot\beta(t-1)} \tilde h_{\varphi,}{}^{\dot\psi}]
\end{multline*}
and so on. As a result,
\begin{multline}
I^{\alpha(t-1)}{}_{,}{}^{\dot\beta(t-1)} =
- a_1 \tilde D \Psi^{\alpha(t-1)}{}_{,}{}^{\dot\beta(t-1)} (2, 0) -\\
- a_2 \tilde D \Psi^{\alpha(t-1)}{}_{,}{}^{\dot\beta(t-1)} (3, 0) - \ldots
- a_n \tilde D \Psi^{\alpha(t-1)}{}_{,}{}^{\dot\beta(t-1)} (t-1, 0) \simeq \\
\simeq \Phi_{1}{}^{\alpha(t-1)}{}_{,}{}^{\dot\beta(t-1)} - \Phi_{2}{}^{\alpha(t-1)}{}_{,}{}^{\dot\beta(t-1)}
\end{multline}
and
$I^{\alpha(t-1)}{}_{,}{}^{\dot\beta(t-1)}
\neq J^{\alpha(t-1)}{}_{,}{}^{\dot\beta(t-1)}$
for $c_{ij}$ of any symmetry.

\end{document}